\documentclass[prb,amsmath,amssymb,superscriptaddress,showpacs,floatfix]{revtex4}
\usepackage{graphicx}% Include figure files
\usepackage{dcolumn}% Align table columns on decimal point
\usepackage{bm}% bold math
\begin{document}

\title{Surface spin dynamics of antiferromagnetically \\ coupled frustrated triangular films}

\author{E. Meloche}
\email{eric@physics.mun.ca}
\affiliation{Department of Physics and Physical Oceanography, Memorial University of Newfoundland, St John's, Newfoundland, Canada A1B 3X7}

\author{M. L. Plumer}
\affiliation{Department of Physics and Physical Oceanography, Memorial University of Newfoundland, St John's, Newfoundland, Canada A1B 3X7}

\author{C. M. Pinciuc}
\affiliation{Edward S. Rogers Sr. Department of Electrical and Computer
Engineering, University of Toronto, Ontario, Canada M5S 3G4}

\date{\today}% It is always \today, today,
             %  but any date may be explicitly specified

\begin{abstract}
Results are presented for spin-wave dispersions in geometrically frustrated stacked triangular antiferromagnets with a thin film or semi-infinite geometry having either zero, easy-plane, or easy-axis anisotropy.    Surface effects on the equilibrium spin configurations and excitation spectrum are investigated for the case of antiferromagnetically coupled films, serving to extend previous results on ferromagnetically coupled layers [E. Meloche \textit{et al.}, Phys. Rev. B \textbf{74}, 094424 (2006)].  An operator equation of motion formalism  is applied to systems which are quasi-one and quasi-two dimensional in character.   In contrast to the case of ferromagnetically coupled films the new results show surface modes that are well separated in frequency from bulk excitations.   Magnetic excitations in thin films with an even or an odd number of  layers show qualitatively different behavior.  These results are relevant for a wide variety of stacked triangular antiferromagnetics materials.
\end{abstract}

\maketitle

\section{Introduction}
   
   The significant amount of research effort devoted to the discovery and understanding of new aspects of geometrically frustrated magnetic systems in recent decades has been almost exclusively focused on the study of bulk properties.\cite{Moessner1}  Among such systems, a large number of materials have been identified which are realizations of the prototype frustrated triangular antiferromagnet (AF).  In addition to the AF near-neighbor in-plane coupling, the vast majority of these systems are also characterized by AF coupling between stacked triangular layers, such as in most of the ABX$_3$ compounds.\cite{Collins1}   Theoretical and experimental studies of magnetic excitations in these model materials have revealed important information on the nature of the fundamental interactions and exposed new types of spin-wave modes in the bulk systems.\cite{Oyedele1,Nagler1,Kimura1,Laar1}   Interest in surface effects on frustrated magnetic systems has been enhanced recently with the fabrication of new materials classes. Studies of hexagonal magnetoelectric and multiferroic rare-earth magnetites (e.g., RMnO$_3$) thin films are being fueled by potential technological applications in the field of spintronics.\cite{Fiebig1}  Ultrathin films of Mn on fcc (111) Fe-Ni substrates also offer the possibility to study AF triangular layers.\cite{Li1}   The understanding of thin-film, surface and interface spin dynamics of such systems is of interest for the engineering of high data transfer rate devices.\cite{Plumer1}    
   
     In the present work, results are shown of solving the operator equation of motion to examine surface spin excitations in a model of AF coupled frustrated triangular layers which have either easy-plane or easy-axis anisotropy.  This study represents an extension of our previous work where the technique was developed and applied to the case ferromagnetically (F) coupled triangular layers having easy-plane anisotropy [E. Meloche \textit{et al.}, Phys. Rev. B \textbf{74}, 094424 (2006)].  (A summary of general theoretical and experimental techniques that have been employed to study thin-film and surface spin waves in un-frustrated systems is also provided in Ref. [10].)   The formulation of a linear spin wave theory is based on a model Hamiltonian which includes isotropic and anisotropic exchange as well as single-ion anisotropy and allows for layer-dependent coefficients of these interactions. Depending on the specific system of interest (e.g., semi-infinite and thin films with or without adjacent layers of a different material) surface parameters may have values quite different from the bulk. A significant difference found for the present case of AF-coupled layers is that localized modes can be well separated in frequency even with bulk parameter values used at the surface layers, in contrast with F-coupled films. Also, as might be expected, equilibrium spin configurations and excitations are dependent on whether the films are composed of an even or odd number of layers (again in contrast with F-coupled films). Due to physical geometrical considerations, nearly all material realizations of stacked triangular layers can be classified as having either quasi-one or quasi-two dimensional magnetic character.    Illustrative numerical results of the spin wave calculations for these cases are discussed in the following sections with the intent to reveal key aspects of thin-film and surface effects on spin excitations in these systems.   

The systems considered here may be represented using the following spin Hamiltonian 

\begin{eqnarray}\label{Hamiltonian}
\mathcal{H}=\sum_{<i,j>}J_{i,j}\Big(\boldsymbol{S}_i\cdot
\boldsymbol{S}_j+\sigma S_i^z
S_j^z\Big)+\sum_{<i,i'>}J'_{i,i'}\Big(\boldsymbol{S}_i\cdot\boldsymbol
{S}_{i'}+\sigma' S_i^z S_{i'}^z\Big)+\sum_{i}D_i(S_i^z)^2
\end{eqnarray}

\noindent where $J_{i,j}>0$ represents the intralayer nearest-neighbor AF exchange coupling and $J'_{i,i'}>0$ represents the interlayer AF exchange coupling.  The effects of anisotropic exchange coupling are included through the parameters $\sigma$ and $\sigma'$, and $D_i$ represents the strength of the single-ion anisotropy at a magnetic site labelled $i$.  The Hamiltonian in Eq. (\ref{Hamiltonian}) can used to describe systems characterized with easy-plane ($D_i>0$) and easy-axis ($D_i<0$) single-ion anisotropy.   Easy-plane and easy axis systems will have different equilibrium spin configurations and spin dynamics and therefore the cases will be treated separately.   
In Sec. \ref{Ham2}, bulk and surface spin-wave modes are calculated for the case of easy-plane anisotropy.  In Sec. \ref{easec}  results are deduced for the spin-wave energies and spectral intensities for the case of easy-axis anisotropy.  A discussion of these results and their relevance to specific compounds is presented in Sec. \ref{Conclusions}.

%%%%%%%%%%%%%%%%%%%%%%%%%%%%%%%%%%%%%%%%%%%%%%%%%%%%%%%%%%%%%%%%%%%%%%%%%%%%%%%%%%%%%%%%%%%%%%%%%%%%%%%%%%%55
\section{easy-plane anisotropy}\label{Ham2}

\subsection{Easy-plane bulk spin-wave excitations}\label{bulksw}

Extending the general method described in Ref. [11] to calculate the bulk spin-wave dispersion relation, the system is divided into three pairs of sublattices on two adjacent layers, which are labeled $A_1$, $B_1$, $C_1$; $A_2$, $B_2$, and $C_2$, and the specific sites on each sublattice are labeled with indices $l_1$, $m_1$, $n_1$, $l_2$, $m_2$, and $n_2$, respectively.  Within any particular layer the $120^\circ$ structure is stabilized  as a result of the pairwise antiferromagnetic coupling between nearest-neighbor spins.  Also, in the classical ground state configuration the spins on adjacent layers (along the crystallographic c axis) order antiparallel to one another.   We first transform to a local coordinate system such that the $z$-axis for each sublattice is in the direction of the average spin alignment.   The transformations to the local coordinate systems for the various sites may be written as

\begin{eqnarray}\label{transformation}
(S_{i}^x,S_{i}^y,S_{i}^z)&\rightarrow &\big[(-1)^{\mu+1}(S_{i}^{x'}\cos\theta_i -S_{i}^{z'}\sin\theta_i) ,(-1)^{\mu+1}(S_{i}^{x'}\sin\theta_i +S_{i}^{z'}\cos\theta_i) ,-S_{i}^{y'}\big]
\end{eqnarray}

\noindent with  $i=l_\mu,m_\mu n_\mu$ and  $\theta_{l_\mu}=0$, $\theta_{m_\mu}=2\pi/3$, $\theta_{n_\mu}=-2\pi/3$ with $\mu=1,2$.  The linearized equations of motion for the local spin raising and lowering operators at each lattice site are formed and then transformed to a wavevector representation. Assuming the usual time dependence $\exp(-iEt)$, the resulting set of 12 equations may be expressed as $\mathbf{Mb}=0$, where

\begin{eqnarray}\label{matrixM}
\mathbf{M}=\left(\begin{array}{cc}
\mathbf{A}&\boldsymbol{\tau}\\
\boldsymbol{\tau}&\mathbf{A}\\
\end{array}\right)\quad\textrm{and}\quad\mathbf{b}=\left[\begin{array}{cc}
\mathbf{b_1}\\\mathbf{b_2}
\end{array}\right].
\end{eqnarray}

\noindent The elements of the $6\times6$ block-circulant matrices $\mathbf{A}$ and
$\boldsymbol{\tau}$ are written as

\begin{eqnarray}\label{A1}
\mathbf{A}=\left(\begin{array}{ccc}
\tilde{A}&\tilde{B}&\tilde{B}^*\\
\tilde{B}^*&\tilde{A}&\tilde{B}\\
\tilde{B}&\tilde{B}^*&\tilde{A}\\
\end{array}\right)\quad \textrm{,}\quad\boldsymbol{\tau}=\left(\begin{array}{ccc}
{\tau}(k_z)&0&0\\
0&{\tau}(k_z)&0\\
0&0&{\tau}(k_z)\\
\end{array}\right).
\end{eqnarray}

\noindent The matrix elements of $\mathbf{A}$ and $\boldsymbol{\tau}$ depend on the energy and system parameters and are defined as

\begin{eqnarray}\label{A2}
\tilde{A}=\left(\begin{array}{cc}
E+\Omega&\alpha\\
-\alpha&E-\Omega
\end{array}\right)\quad \textrm{,}\quad \tilde{B}=\left(\begin{array}{cc}
\beta & \gamma\\
-\gamma & -\beta
\end{array}\right)\quad \textrm{and}\quad {\tau}(k_z)=SJ'(k_z)/2\left(\begin{array}{cc}
-\sigma' &2+\sigma'\\
-2-\sigma' &\sigma'\end{array}\right)
\end{eqnarray}
\noindent where
\begin{eqnarray}\label{p1}
\Omega&=&-SJ(0)-SD'-SJ'(0)\nonumber\\
\alpha&=&SD'\nonumber\\
\beta&=&-S(1/2+\sigma)J(\mathbf{k}_\parallel)/2\nonumber\\
\gamma&=&S(3/2+\sigma)J(\mathbf{k}_\parallel)/2\nonumber\\
D'&=&[1-(2S)^{-1}]D.
\end{eqnarray}

The exchange integrals $J(\mathbf{k}_\parallel)$ and $J'(k_z)$ are
defined as

\begin{eqnarray}\label{exch1}
J(\mathbf{k}_\parallel)&=&J\big(2\cos(k_y \sqrt{3}a/2)\exp(-i k_x a/2)+\exp(i k_x a)\big)\nonumber\\
J'(k_z)&=&2J'\cos(k_z c)
\end{eqnarray}
\noindent  with $a$ and $c$ denoting the lattice constants.  The elements of the column vector $\mathbf{b}$ are written as

\begin{eqnarray}\label{bn}
\mathbf{b_\mu}=[S_{A_\mu}^{+}(\mathbf{k}),S_{A_\mu}^{-}(-\mathbf{k}),S_{B_\mu}^{+}(\mathbf{k}),S_{B_\mu}^{-}(-\mathbf{k}),S_{C_\mu}^{+}(\mathbf{k}),S_{C_\mu}^{-}(-\mathbf{k})]^T
\end{eqnarray}

\noindent for $\mu=1,2$. The terms $S_{A_\mu}^{\pm}(\mathbf{k})$ correspond to the Fourier amplitudes of the spin operators $S_{l_\mu}^{\pm}$, for $\mu=1,2$, along with similar definitions for $S_{B_\mu}^{\pm}(\mathbf{k})$ and $S_{C_\mu}^{\pm}(\mathbf{k})$.  The bulk spin-wave modes correspond to the solutions of $\det \mathbf{M}=0$.  Analytical results may be obtained by first block diagonalizing $\mathbf{M}$ using the transformation $\mathbf{W}^{\dagger}\mathbf{M}\mathbf{W}=\mathbf{D}$ where

\begin{eqnarray}\label{t1}
\mathbf{W}=\frac{1}{\sqrt{2}}\left(\begin{array}{cc}
\mathbf{U}&\mathbf{U}\\
-\mathbf{U}&\mathbf{U}\\
\end{array}\right),
\end{eqnarray}

\begin{eqnarray}\label{t2}
\mathbf{U}=\frac{1}{\sqrt{3}}\left(\begin{array}{ccc}
\openone_2&\openone_2&\openone_2\\
\openone_2&x\openone_2&x^*\openone_2\\
\openone_2&x^*\openone_2&x\openone_2\\
\end{array}\right)\textrm{with}\quad x=\exp(i2\pi/3)
\end{eqnarray}

\noindent  The nonzero elements of $\mathbf{D}$ consist of six $2\times2$ matrices along the main diagonal and the bulk spin-wave energies are obtained from the solutions of the determinantal conditions $\det(\tilde{\Lambda}(\phi)\pm {\tau})=0$ for $\phi=0,\pm 2\pi/3$ where we have defined $\tilde{\Lambda}(\phi)=\tilde{A}+\tilde{B}\exp(i\phi)+\tilde{B}^*\exp(-i\phi)$.  Bulk spin-wave energies may be expressed as $\omega_{\mathbf{k\pm}}(\phi)$ with $\phi=0,\pm 2\pi/3$, where

\begin{eqnarray}\label{bulkE}
\omega_{\mathbf{k}\pm}^{2}(\phi)=\big(\Omega +\alpha\pm SJ'(k_z)+SJ\Phi(\mathbf{k}_\parallel ,\phi) \big) \big(\Omega
-\alpha\mp S(1+\sigma' )J'(k_z)-2SJ(1+\sigma)\Phi(\mathbf{k}_\parallel ,\phi)\big)
\end{eqnarray}

\noindent and

\begin{eqnarray}\label{bulkphi}
\Phi(\mathbf{k}_\parallel ,\phi)=2\cos(\sqrt{3} k_y a/2) \cos(-k_x
a/2 + \phi) + \cos(k_x a + \phi).
\end{eqnarray}

%%%%%%%%%%%%%%%%%%%%%%%%%%%%%%%%%%%%%%%%%%%%%%%%%%%%%%%%%%%%%%%%%%%%%%%%%%%%%%%%%%%%%%%%%%%%%%%%%%%%%%%%%%%%%%%%%%%
\subsection{Easy-plane surface spin-waves}\label{SI}
%%%%%%%%%%%%%%%%%%%%%%%%%%%%%%%%%%%%%%%%%%%%%%%%%%%%%%%%%%%%%%%%%%%%%%%%%%%%%%%%%%%%%%%%%%%%%%%%%%%%%%%%%%%%%%%%%%%%

As in our previous work,\cite{Meloche1} calculations are first made for a semi-infinite system with a single (001) surface and for a thin film composed of $N$ magnetic layers which are labeled using the layer index $n$ $(=1,...,N)$.  The linearized equations of motion for the local spin raising and lowering operators at all the sites are formed and then transformed to a representation involving a two-dimensional wavevector $\mathbf{k}_{\parallel}$ which runs parallel to the surfaces and a layer index $n$.  The spins located on the layers with an odd layer index $n$ are taken to belong to sublattices $A_1$, $B_1$, $C_1$ whereas those located on layers with an even layer index $n$ belong to sublattices $A_2$, $B_2$, $C_2$. For the $l^{\textrm{th}}$ site on sublattice $A_1$ the wave-like solution for the spin operator is written as

\begin{eqnarray}
S_{l_1}^\pm=S_{A_1,n}^\pm(\mathbf{k}_{\parallel})\exp[i(\mathbf{k}_{\parallel}\cdot
\boldsymbol{\rho}-Et)]
\end{eqnarray}

\noindent where the position vector $\boldsymbol{\rho}=(x,y)$ and the amplitudes $S_{A_1,n}^\pm(\mathbf{k}_{\parallel})$ depend on the $z$ coordinate through the layer index $n$.   Similar expressions are defined for sites on sublattices $B_1, C_1, A_2, B_2$ and $C_2$. For a semi-infinite system the set of finite difference equations connecting the Fourier amplitudes may be expressed in supermatrix form as $\mathbf{M}\mathbf{b}=0$, where $\mathbf{M}$ is an $\infty\times\infty$ block-tridiagonal matrix defined as

\begin{eqnarray}\label{semiinf1}
\mathbf{M}=\left(\begin{array}{cccccc}
\mathbf{A}_1&{\tau}(0)/2&0&0&0&\cdots\\
{\tau}(0)/2&\mathbf{A}&{\tau}(0)/2&0&0&\cdots\\
0&{\tau}(0)/2&\mathbf{A}&{\tau}(0)/2&0&\cdots\\
\vdots&\ddots&\ddots&\ddots&\ddots&\ddots
\end{array}\right)
\end{eqnarray}

\noindent where each element $\mathbf{M}$ represents a $6\times 6$ matrix and $\mathbf{A}$ is defined as in (\ref{A1})-(\ref{exch1}).  The elements of the matrix $\mathbf{A}_1$ are defined as in $\mathbf{A}$ but with the substitution $J\rightarrow J_1, D\rightarrow D_1$ and $J'\rightarrow J'/2$.   The differences between $\mathbf{A}_1$ and $\mathbf{A}$ occur because of different exchange and anisotropy parameters for spins in the surface layer as well as the reduced number interlayer neighbors for surface spins. The infinite column vector $\mathbf{b}$ is written as $\mathbf{b}=[\mathbf{b}_1,\mathbf{b}_2,...]^T$ where $\mathbf{b}_n$ is defined in Eq. (\ref{bn}) and $n=1,...,\infty$.    The equations are first partially decoupled by applying the transformation $\mathbf{U}^{-1}\mathbf{M}_{i,j}\mathbf{U}$ to all of the elements of the supermatrix.  The transformed equations may be written into three sets of finite difference equations as

\begin{eqnarray}\label{findiff1}
\tilde{\Lambda}_1(\phi)\mathbf{X}_{1,\phi}+\frac{{\tau}(0)}{2}\mathbf{X}_{2,\phi}&=&0\qquad(n=1)
\end{eqnarray}
\begin{eqnarray}\label{findiff2}
\frac{{\tau}(0)}{2}\mathbf{X}_{2\nu,\phi}+\tilde{\Lambda}(\phi)\mathbf{X}_{2\nu+1,\phi}+\frac{{\tau}(0)}{2}\mathbf{X}_{2\nu+2,\phi}&=&0 \qquad(n=2\nu+1,\nu\geq 1)
\end{eqnarray}
\begin{eqnarray}\label{findiff3}
\frac{{\tau}(0)}{2}\mathbf{X}_{2\nu-1,\phi}+\tilde{\Lambda}(\phi)\mathbf{X}_{2\nu,\phi}+\frac{{\tau}(0)}{2}\mathbf{X}_{2\nu+1,\phi}&=&0 \qquad(n=2\nu,\nu\geq 1)
\end{eqnarray}

\noindent for $\phi=0,\pm 2\pi/3$. The matrix $\tilde{\Lambda}_1(\phi)$ is defined as $\tilde{\Lambda}_1(\phi)=\tilde{A}_1+\tilde{B}_1\exp(i\phi)+\tilde{B}_1^\dagger\exp(-i\phi) $, whereas $\tilde{\Lambda}(\phi)$ is defined as in the bulk case.  The column vectors in Eq. (\ref{findiff1})-(\ref{findiff3}) are written as $\mathbf{X}_{n,\phi}=[X_n^+(\phi),X_n^-(\phi)]^T$ with elements 

\begin{eqnarray}
X_n^\pm(\phi)=\frac{1}{\sqrt{3}}\big(S_{A,n}^{\pm}+S_{B,n}^{\pm}\exp(-i\phi)+S_{C,n}^{\pm}\exp(i\phi)\big).
\end{eqnarray}

\noindent where we omit the wavevector label $\mathbf{k}_{\parallel}$. Column vectors $\mathbf{X}_{n,\phi}$ with an odd (even) layer index $n$ involve linear combinations of the operator amplitudes on sublattice $1$ (2) only.  Equations (\ref{findiff1}) and (\ref{findiff2}) are used to eliminate the  $\mathbf{X}_{n,\phi}$ with $n$ odd from the set of equations represented by (\ref{findiff3}).  The resulting set of relations connecting $\mathbf{X}_{n,\phi}$ with $n$ even can be written in the matrix form 

\begin{eqnarray}\label{even}
\left(\begin{array}{cccccc}
\mathbf{P}_{1,\phi}&\mathbf{Q}_\phi&0&0&0&\cdots\\
\mathbf{Q}_\phi&\mathbf{P}_\phi&\mathbf{Q}_\phi&0&0&\cdots\\
0&\mathbf{Q}\phi&\mathbf{P}_\phi&\mathbf{Q}_\phi&0&\cdots\\
\vdots&\ddots&\ddots&\ddots&\ddots&\cdots
\end{array}\right)\left(\begin{array}{c}
\mathbf{X}_{2,\phi}\\
\mathbf{X}_{4,\phi}\\
\mathbf{X}_{6,\phi}\\
\vdots
\end{array}\right)=0
\end{eqnarray}

where 

\begin{eqnarray}
\mathbf{P}_{1,\phi}&=&\tilde{\Lambda}(\phi)-\frac{1}{4}{\tau}(0)(\tilde{\Lambda}_1^{-1}(\phi)+\tilde{\Lambda}^{-1}(\phi)){\tau}(0)\nonumber\\
\mathbf{P}_{\phi}&=&\tilde{\Lambda}(\phi)-\frac{1}{2}{\tau}(0)\tilde{\Lambda}^{-1}(\phi){\tau}(0)\nonumber\\
\mathbf{Q}_{\phi}&=&-\frac{1}{4}{\tau}(0)\tilde{\Lambda}^{-1}(\phi){\tau}(0)
\end{eqnarray}

\noindent The three sets of equations given in Eq (\ref{even}) (for $\phi=0,\pm 2\pi/3$) couple the amplitudes on sublattice 2 only. The surface spin-wave frequencies can be obtained numerically following an analogous approach employed for ferromagnetically coupled layers.\cite{Meloche1}

For a thin film composed of $N$ layers with (001) surfaces the system of finite difference equations can be expressed in a supermatrix form as $\mathbf{Mb}=0$ as for the semi-infite case (see Eq. \ref{semiinf1}) except $\mathbf{M}$ is now a $6N\times 6N$ block-tridiagonal matrix and the column vector is defined as $\mathbf{b}=[\mathbf{b}_1,...,\mathbf{b}_N]^T$.  The first and the last elements on the main diagonal of the supermatrix $\mathbf{M}$ are written as $A_1$ and $A_N$, respectively.  The elements of the matrix $A_N$ describe the effects of the additional surface of the film and are defined as in $A_1$ for the semi-infinite case except with the appropriate exchange and anisotropy parameters at the surface $n=N$.   The coupled systems of equations in the case of a thin film can be decoupled using the method outlined for a semi-infinite system.   However, cases involving an even  or an odd  number of layers $N$ must be dealt with seperately.   For a film containing an odd number of layers both surfaces belong to the same sublattice whereas for a film an even number of layers the spins in the surface layers belong to difference sublattices.   The spin-wave excitation spectrum in a stacked triangular antiferromagnetic film are obtained numerically by solving the determinantal condition $\det\mathbf{M}=0$.  

\subsection{Numerical Results}

\begin{figure}
\begin{center}
\includegraphics[width=0.5\textwidth]{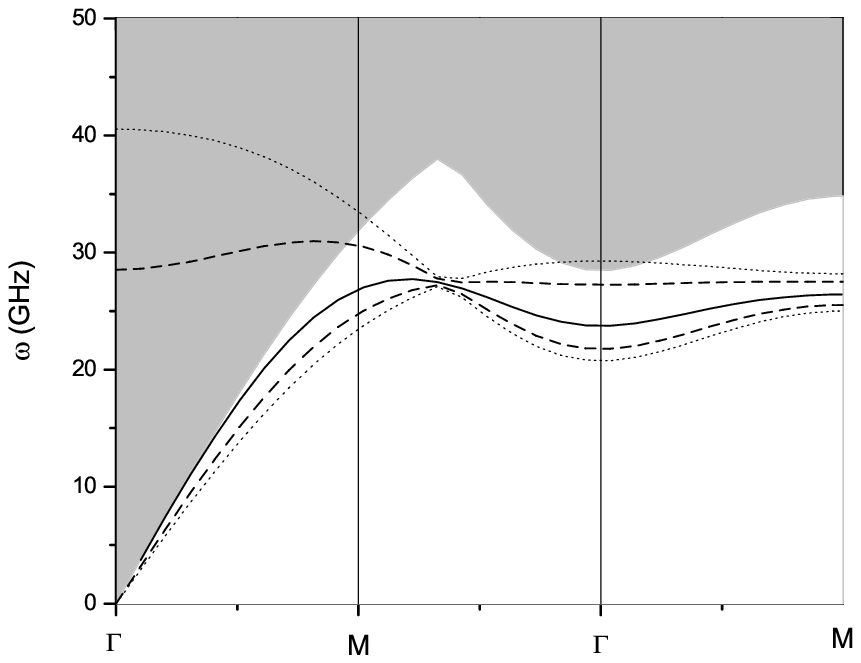}
\includegraphics[width=0.5\textwidth]{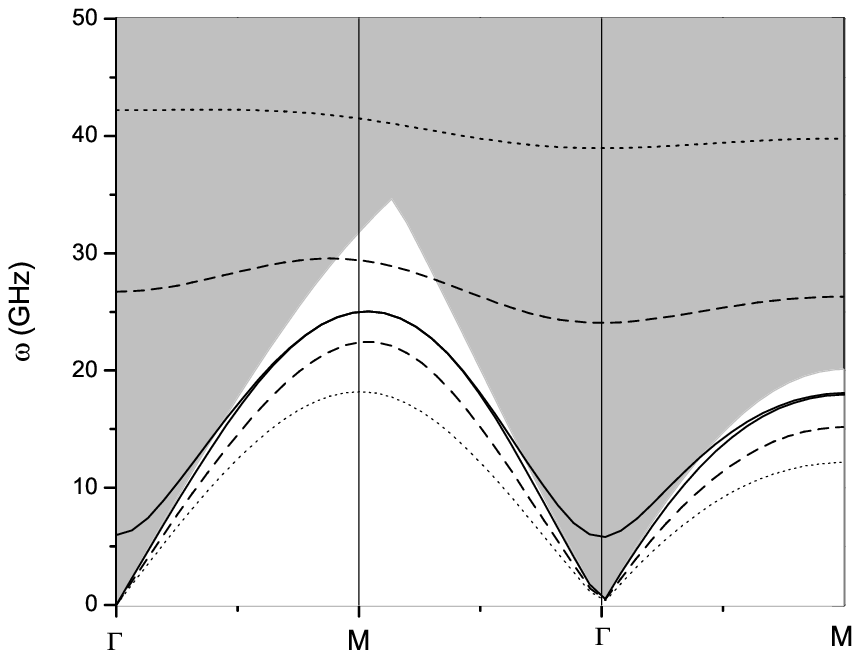}
\caption{\label{Q1DEP} Spin-wave energy versus in-plane wavevector $k_x a$ for a quasi-1D system with  $J=1.0$ GHz, $J'=100.0$ GHz and $\sigma=\sigma'=0.0$ and in the top figure $D'=1.0$ GHz whereas in the bottom $D'=0.0$.  The $M$-point refers to the wavevector $\mathbf{k_\parallel}=(2\pi/3,0)$.  Shaded regions correspond to the bulk excitations.  In the top figure, solid, dotted and dashed curves correspond to the semi-infinite, 8-layer and 16 layers systems, respectively.  In the bottom figure, dotted, dashed and solid lines correspond to films with 9,17 and 100 layers, respectively. } 
\end{center}
\end{figure}

In Fig. \ref{Q1DEP} illustrative results are shown for the spin-wave frequency versus in-plane wavevector $k_x a$ (with $k_y=0$) for a representative $S=1$ quasi-1D planar triangular antiferromagnet where we set  $J=1.0$ GHz, $J'=100.0$ GHz and $\sigma=\sigma'=0.0$.  The dispersion relations in the top and bottom figures correspond to $D'=1.0$ GHz and $D'=0$, respectively.  The surface exchange and anisotropy parameters are taken to be uniform throughout the system.  The shaded areas represent all of the bulk solutions $\omega_{\mathbf{k}\pm}(\phi)$ (see Eq. (\ref{bulkE})) with $0<k_z c\le\pi/2$ and the three sections (from left to right) correspond to the bulk solutions with $\phi=0,2\pi/3,-2\pi/3$. The $M$-point denotes the zone-edge wavevector $(k_x,k_y)=(2\pi/3a,0)$.  For these parameter values the top of the bulk regions extends to approximately 208 GHz and only the lower part of the effective bulk continuum is shown.   The solid line represents a surface spin-wave obtained from the solution of Eq. (\ref{even}) for a semi-infinite system. This new mode is characterized with an amplitude that decays into the film away from the surface.   In the top figure (zero anisotropy) results are also shown for two lowest energy modes for thin films composed layers 8 (dotted lines) and 16 layers (dashed lines).  In the bottom figure the dotted, dashed and solid lines represents two lowest energy excitations for a film composed $N=$9, 17 and 100 layers, respectively.   The dominant effect of the anisotropy is to open up a gap in the spectrum at the $\Gamma$-point for the surface and bulk excitation. The gap energies for the bulk modes are obtained from Eq. (\ref{bulkE}) and are $\omega_{\mathbf{k}-}(\pm 2\pi/3)=S[D'(8J'+9J)]^{1/2}$.    For most values of the wavevector the surface mode obtained for a semi-infinite system is well separated from the bulk region.  This contrasts with the behavior obtained for a quasi-1D easy-plane triangular antiferromagnet with \textit{ferromagnetic} interlayer exchange coupling where the splitting between the surface branch and the bulk region is negligible for a system characterized with uniform exchange and anisotropy parameters.   

 The spin-wave dispersion relation in these frustrated films show interesting characteristics.  The dispersion relation in films composed of odd number of layers are qualitatively similar to those with \textit{ferromagnetic} interlayer exchange.   This can be explained by the fact that for an odd number of layers both surfaces belong to the same sublattice and each individual chain of spins will possess a net moment.   For thin films with an even number of layers $N$ the system is composed of antiferromagnetically coupled chains that possess no net moment because each chain of spins has an equal number of sites from sublattices 1 and 2.  As the number of layers in the film increases this effect becomes less important and the two lowest branches eventually become degenerate and equal to the results obtained for the semi-infinite system.  The two modes correspond to localized excitations at each surface.

\begin{figure}
\begin{center}
\includegraphics[width=0.5\textwidth]{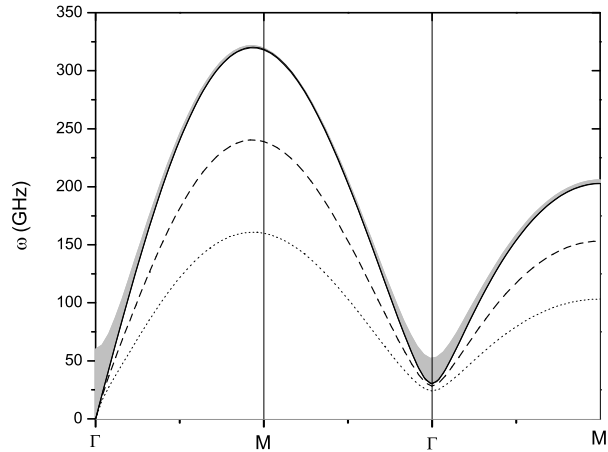}
\caption{\label{Q2DEP} Surface and bulk spin-wave energy versus in-plane wavevector $k_x a$ for a semi-infinite quasi-2D system with $J=100.0$ GHz, $J'=1.0$ GHz, and $D'=1.0$ GHz (with $D_1=D$) and $\sigma=\sigma'=0.0$. The solid line appearing just below the bulk region (shaded area) corresponds to a case with uniform interlayer coupling $J_1=J$, whereas the dotted and dashed lines are obtained with $J_1=0.5J$ and $J_1=0.75J$, respectively. } 
\end{center}
\end{figure}

In Fig. \ref{Q2DEP} results are shown for the surface and bulk spin-wave energies versus in-plane wavevector $k_x a$ (with $k_y=0$) for a semi-infinite quasi-2D system with $J=100.0$ GHz, $J'=1.0$ GHz, $D'=1.0$ GHz (with $D_1=D$) and $\sigma=\sigma'=0.0$.   The bulk spin waves form a very narrow continuum (shaded areas) in quasi-2D systems because of the weak dependence on the third wavevector component $k_z$.   Also shown are the substantial effects of a modified surface exchange coupling $J_1$ on the surface spin waves.     Splitting of the surface branch away from the bulk region can also be obtained by assuming a modified value for the surface anisotropy $D_1$ compared to the bulk value.  The energy gap at the zone center $\Gamma$ for the bulk modes $\omega_{\mathbf{k}-}(\pm 2\pi/3)$ is due to the anisotropy and vanishes when $D'=0$.

\section{easy-axis anisotropy}\label{easec}

\subsection{Easy-axis bulk spin-wave excitations}\label{bulkswea}

\begin{figure}
\begin{center}
\includegraphics[angle=270,width=0.8\textwidth]{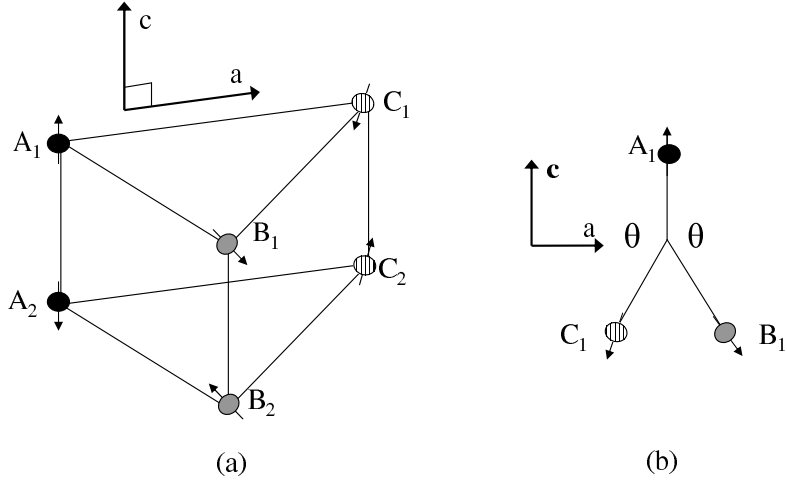}
\caption{\label{eabulk2} (a)  The deformed 120$^\circ$ structure for a STA with easy-axis anisotropy.  The angle $\theta$ is greater than $2\pi/3$ when $D<0$ (when $\sigma=\sigma'=0.0$).   All of the spin lie in the $ac$ plane and the direction of the spins on sublattices $A_2$, $B_2$, $C_2$ are opposite to those on sublattice $A_1$, $B_1$, $C_1$, respectively.}
\end{center}
\end{figure}

Consider now the triangular antiferromagnet characterized with a single-ion anisotropy whose easy-axis is along the crystallographic $c$ axis.   It is instructive to first carry out the calculations for the bulk spin-wave dispersion. We follow a method similar to the one described previously for the case of easy-plane anisotropy.    For the bulk spin-wave calculation, the system can again be divided into the six sublattices as depicted schematically in Fig. \ref{eabulk2}a.   It is worth noting that the ground state spin configuration can be rotated about the $c$ axis by any arbitrary angle and here we assume that all of the spins have their equilibrium positions in the $ac$ plane.  Without loss of generality we set the equilibrium direction for spins on sublattice $A_1$ along the $c$ axis.  The equilibrium configurations for spins on sublattices $B_1, C_1, A_2,B_2,C_2$ can be obtained using a local field method.\cite{Suzuki1,Watabe1}   The main effect of the easy-axis anisotropy on the equilibrium configuration is that the spins now lie in the a plane containing  the $c$ axis and a deformation of the 120$^\circ$ structure giving $\theta > 120^\circ$ as shown in Fig. \ref{eabulk2}b.   

The Hamiltonian in Eq. (\ref{Hamiltonian}) is transformed to a local coordinate system such that the direction of the local $z$-axis is in the direction of the equilibrium spin alignment.  The transformations for the sites on the six sublattices are written as

\begin{eqnarray}
(S_{i}^x,S_{i}^y,S_{i}^z)&\rightarrow &\big[(-1)^{\mu+1}(S_{i}^{x'}\cos\theta_i - S_{i}^{z'}\sin\theta_i) , S_{i}^{y'},(-1)^{\mu+1}( S_{i}^{x'}\sin\theta_i +S_{i}^{z'}\cos\theta_i)\big]
\end{eqnarray}

\noindent with $i=l_\mu,m_\mu,n_\mu$ and  $\theta_{l_\mu}=0$, $\theta_{m_\mu}=\theta$, $\theta_{n_\mu}=-\theta$ with $\mu=1,2$.  The stability condition requires that the coefficients of the local transverse spin components $S_{i}^{x'}$ and $S_{i}^{y'}$ vanish for all sites and leads to the bulk canting angle which is given by

\begin{eqnarray}\label{cantbulk}
\cos\theta=\frac{-(1+\sigma)}{2[1+\sigma/2-\sigma'J'/3J+D'/3J]}.
\end{eqnarray}

\noindent   Eq. (\ref{cantbulk}) is a generalization of the results obtained in Refs. [13,14] which treat the cases of either easy-axis anisotropy or exchange anisotropy, but not both.   In the absence of anisotropy ($\sigma=\sigma'=D'=0$) the 120$^\circ$ structure is recovered, as expected.  The linearized equations of motion for the various sites are calculated and then transformed to a wavevector representation.  Twelve equations are required to obtained a closed set.   The system of coupled equations may be expressed as $\mathbf{Mb}=0$ where the $12\times 12$ matrix $\mathbf{M}$ has the same form as in Eq. (\ref{matrixM}) and the column vector $\mathbf{b}$ is formally written as in Eq. (\ref{bn}) .  In this case, the elements of the supermatrix $\mathbf{M}$ are now given by

\begin{eqnarray}\label{bulkeamat}
\mathbf{A}=\left(\begin{array}{ccc}
\tilde{A}&\tilde{B}&\tilde{B}^*\\
\tilde{B}^*&\tilde{C}&\tilde{D}\\
\tilde{B}&\tilde{D}^*&\tilde{C}\\
\end{array}\right)\quad \textrm{,}\quad\boldsymbol{\tau}=\left(\begin{array}{ccc}
{\tau}(k_z)&0&0\\
0&{\lambda}(k_z)&0\\
0&0&{\lambda}(k_z)\\
\end{array}\right)
\end{eqnarray}
\noindent where
\begin{eqnarray}
\tilde{A}&=&\left(\begin{array}{cc}
E+\Omega&0\\
0&E-\Omega
\end{array}\right);\quad \tilde{B}=-SJ(\mathbf{k}_\parallel)/2\left(\begin{array}{rr}
 c_1^+&c_1^- \\
-c_1^- & -c_1^+
\end{array}\right)\nonumber\\ \tilde{C}&=&\left(\begin{array}{cc}
E+ \alpha& \delta\\
-\delta & E-\alpha
\end{array}\right);\quad \tilde{D}=-SJ(\mathbf{k}_\parallel)/2\left(\begin{array}{rr}
 c_2^+&c_2^- \\
-c_2^- & -c_2^+
\end{array}\right)\nonumber\\ {\tau}(k_z)&=&SJ'(k_z)\left(\begin{array}{rr}
0 &1\\
-1&0\end{array}\right);\quad {\lambda}(k_z)=SJ'(k_z)/2\left(\begin{array}{cc}
\sigma'\sin^2\theta &(2+\sigma'\sin^2\theta)\\
-(2+\sigma'\sin^2\theta)&-\sigma'\sin^2\theta\end{array}\right)\nonumber\\
\end{eqnarray}

\noindent with matrix elements defined as
\begin{eqnarray}\label{bulkele}
\Omega&=&2S(1+\sigma)\cos\theta J(0)+2SD'-S(1+\sigma')J'(0)\nonumber\\
\alpha&=&S((1+\sigma)(\cos\theta+\cos^2\theta)-\sin^2\theta)J(0)-S(1-3\cos^2\theta)D'-S(1+\sigma'\cos^2\theta)J'(0)\nonumber\\
\delta&=&-S D'\sin^2\theta\nonumber\\
c_1^\pm&=&\cos\theta\pm1\nonumber\\
c_2^\pm&=&\cos^2\theta-(1+\sigma)\sin^2\theta\pm 1\nonumber\\
D'&=&[1-(2S)^{-1}]D.
\end{eqnarray}

The exchange integrals $J(\mathbf{k}_\parallel)$ and $J'(k_z)$ are defined as in Eq. (\ref{exch1}).  The matrix $\mathbf{A}$ is not a block-circulant matrix due to the effects of the anisotropy and analytical solutions for the bulk spin-waves cannot be obtained using the diagonalization procedure employed in Eq. (\ref{t1})-(\ref{t2}).   Instead, the bulk spin-wave spectrum is obtained numerically by solving the determinantal condition $\det\mathbf{M}=0$.   

\subsection{Spin-waves in thin films }

\begin{figure}
\begin{center}
\includegraphics[angle=270,width=0.8\textwidth]{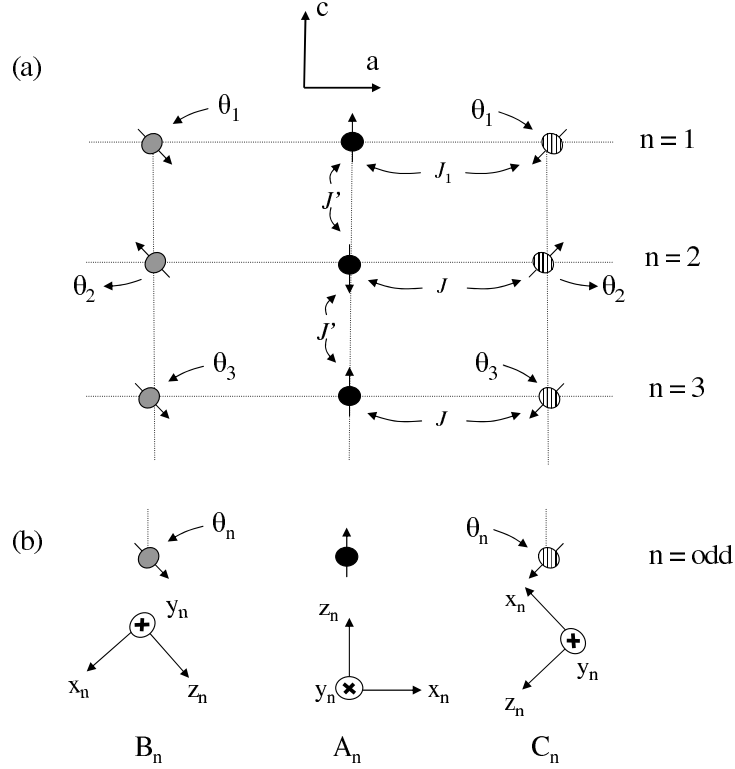}
\caption{\label{newea} (a) Schematic view of the spin configuration for a stacked triangular antiferromagnet with easy-axis anisotropy and a $(001)$ surface as viewed along the chains.  (b) Local coordinate system for spins in layer $n$, where $n$ is an odd layer index. }
\end{center}
\end{figure}

The spin-wave excitations in thin films composed of $N$ magnetic layers with a (001) surfaces may be investigated by extending the method used for bulk spin-waves.  The calculations for a thin film with easy-axis anisotropy are algebraically more complicated than those for a film with easy-plane anisotropy because the system can no longer be described in terms of a six sublattice model.   However, within any particular layer of the film the system is still divided into three interpenetrating sublattices.   Without loss of generality, the equilibrium direction for the surface layer spins $n=1$ on sublattice $A$ is set along the $c$ axis and the equilibrium direction for all of the other spins are obtained by miminizing the classical energy of the system.    Sites on sublattice $A$ with an odd layer index $n$ also have their equilibrium directions along the $c$ axis, whereas sublattice $A$ sites with an even layer index are antiparallel because of the antiferromagnetic interlayer exchange coupling.  The canting angles for sites on sublattices $B$ and $C$ in layer $n$ are written as $\theta_{B,n}=\theta_{C,n}=\theta_{n}$.     A schematic diagram of the system is illustrated in Fig. \ref{newea}a along with the local set of axes for each sublattice.    The $3N$ coordinate transformations may be written as 

\begin{eqnarray}
(S_{i}^x,S_{i}^y,S_{i}^z)&\rightarrow &\big[(-1)^{n+1}(S_{i}^{x'}\cos\theta_i - S_{i}^{z'}\sin\theta_i) , S_{i}^{y'},(-1)^{n+1}( S_{i}^{x'}\sin\theta_i +S_{i}^{z'}\cos\theta_i)\big]
\end{eqnarray}

\noindent for $i=l_n,m_n,n_n$ with $n=1,...,N$, and we define $\theta_{l_n}=0$, $\theta_{m_n}=\theta_n$, $\theta_{n_n}=-\theta_n$.  The Hamiltonian in Eq. (\ref{Hamiltonian}) is again written with respect to the local coordinate systems and the stability conditions for the spins on sublattices $B$ (or $C$) are now given by

\begin{eqnarray}\label{canting}
3J_n\big[(1+\sigma)\sin\theta_n+(1+\sigma/2)\sin 2\theta_n\big]+D'_n\sin 2\theta_n\nonumber\\
+J'\big[(\sin\theta_{n-1}\cos\theta_n-(1+\sigma')\cos\theta_{n-1}\sin\theta_n)(1-\delta_{1,n})\nonumber\\
(\sin\theta_{n+1}\cos\theta_n-(1+\sigma')\cos\theta_{n+1}\sin\theta_n)(1-\delta_{n,N})\big]=0\nonumber\\ \quad \textrm{for} \quad n=1,...,N.
\end{eqnarray}

The surface effects are taken into account through the Kronecker $\delta$ in Eq. ({\ref{canting}}) and modified values of the exchange and single-ion anisotropy at the surfaces of the film are allowed.   The set of $N$ nonlinear equations can be solved using a numerical self-consistent approach.\cite{Nortemann1,Wang1}  However, there are a few limiting cases that can be solved analytically.   If we neglect the anisotropy terms $(\sigma=\sigma'=D'_n=0)$ we recover the $120^\circ$ structure in any particular layer, as expected.  In the limit of $J'=0$ the layers are independent from one another and the canting angle within any layer will be given by Eq. (\ref{cantbulk}).  When the exchange and anisotropy surface parameters are set equal to their bulk values it is easily verified that the canting angles for sites on sublattices $B$ and $C$ are layer independent and again given by Eq. (\ref{cantbulk}).  For all other cases, a numerical solution is required.

The spin-wave energies in thin films are obtained by forming $6N$ linearized coupled equations of motion which may be expressed as $\mathbf{Mb}=0$ where $\mathbf{M}$ is a block tridiagonal matrix defined as

\begin{eqnarray}\label{eqmoeatf}
\mathbf{M}=\left(\begin{array}{cccccc}
\mathbf{A}_1&\boldsymbol{\tau_{1,2}}&0&0&\cdots&0\\
\boldsymbol{\tau_{1,2}}&\mathbf{A}_2&\boldsymbol{\tau_{2,3}}&0&\cdots&\vdots\\
\vdots&\ddots&\ddots&\ddots&\cdots&0\\
0&0&0&\boldsymbol{\tau_{N-2,N-1}}&\mathbf{A}_{N-1}&\boldsymbol{\tau_{N-1,N}}\\
0&0&\cdots&0&\boldsymbol{\tau_{N-1,N}}&\mathbf{A}_N
\end{array}\right)
\end{eqnarray}

\noindent and $\mathbf{b}=[\mathbf{b}_1,...,\mathbf{b}_N]^T$.  The elements of the matrices $\mathbf{A}_n$ and $\boldsymbol{\tau}_{n,n'}$ (defined in Appendix A) are similar to those in the bulk case in Eqs. (\ref{bulkeamat})-(\ref{bulkele}) where the main differences are due to the layer dependent canting angles for spins on sublattices $B$ and $C$.  Spin-wave excitations are obtained numerically from the solutions of $\det\mathbf{M}=0$.

\subsection{Numerical Results}

\begin{figure}
\begin{center}
\includegraphics[width=0.5\textwidth]{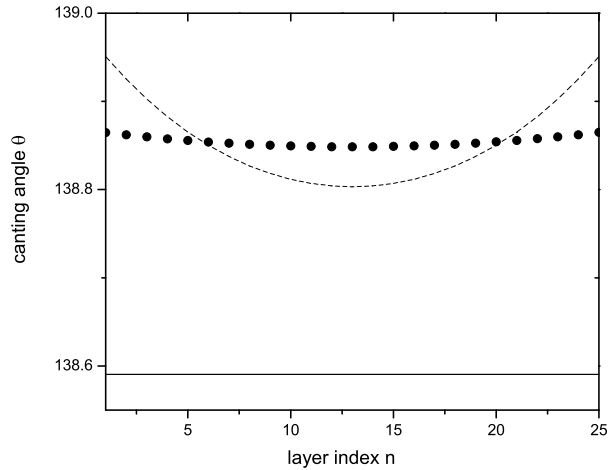}
\caption{\label{cant1} Canting angle (measured in degrees) for sites on sublattices $B$ or $C$ versus layer index for a $N=25$ layers film with easy-axis anisotropy.  All the curves correspond to $D'_1=D'=D'_N=-J$ and $\sigma=\sigma'=0$.  The solid line corresponds to the case $J_1=J=J_N$ and the canting angle (Eq. (\ref{cantbulk})) is independent of the coupling strength $J'$.   The dashed line and the circles show results for the quasi-1D cases of $J'=100 J$ and $J'=1000 J$ when the surface exchange couplings differ slightly from the bulk values ($J_1=J_N=0.9J$).} 
\end{center}
\end{figure}

In Fig. \ref{cant1} results are shown for the canting angle for spins occupying sites on sublattice $B$ or $C$ in a  25-layer film with easy-axis anisotropy.   In all cases the spins on sublattice $A$ form antiferromagnetic chains with ordering along the $c$ axis.  The canting angle is layer dependent only if the surface exchange or anisotropy parameters are allowed to differ from their corresponding bulk values.  In quasi-1D systems the strong interlayer coupling tends to restore the antiferromagnetic order in the spin chains on sublattices $B$ and $C$.    For quasi-2D systems the perturbation of the canting angle in cases with modified surface parameters is localized to a small number of layers (typically $\sim 2$) away from the surface.

\begin{figure}
\begin{center}
\includegraphics[width=0.5\textwidth]{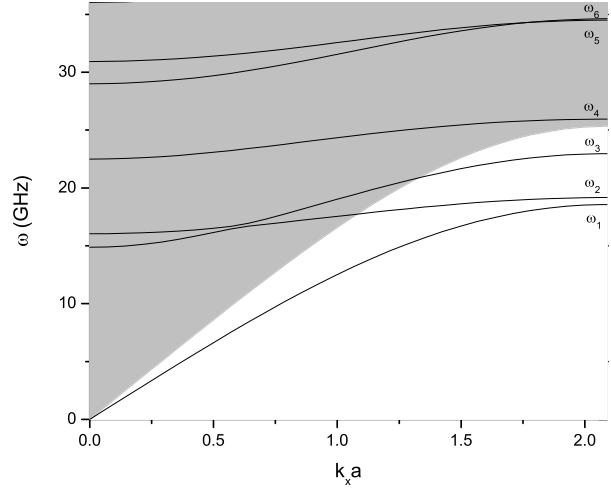}
\caption{\label{Q1DEA} Spin-wave energy versus in-plane wavevector $k_x a$ (with $k_y=0$) for a $S=1$ quasi-1D system with easy-axis anisotropy.  The parameters are $J=100.0$ GHz, $J'=1.0$ GHz, $D'=-1.0$ GHz, $\sigma=\sigma'=0.0$. The exchange and anisotropy parameters at the surfaces of the film are set equal to their corresponding bulk values.  The shaded area corresponds to the lower edge of the bulk continuum and the solid lines correspond the six lowest energy modes obtained for a thin film composed of $N=20$ layers.  The six lowest energy branches for the film are labelled $\omega_1,...,\omega_6$.} 
\end{center}
\end{figure}

An example of the spin-wave dispersion relation for a quasi-1D system with easy-axis anisotropy is illustrated in Fig. \ref{Q1DEA}.   The shaded area corresponds to the low-energy bulk spin-waves. The solid lines represent the $six$ lowest energy modes for a uniform film composed of $N=20$ layers.   The parameter values for the representative quasi-1D system are $J=1.0$ GHz, $J'=100.0$ GHz, $D'=-1.0$ GHz and $\sigma=\sigma'=0.0$.  All of the spin-waves modes obtained from the determinental condition $\det\mathbf{M}=0$ are degenerate in magnitude and only the positive solutions are shown.

\begin{figure}
\centering
\begin{tabular}{cc}
\includegraphics[width=0.48\textwidth]{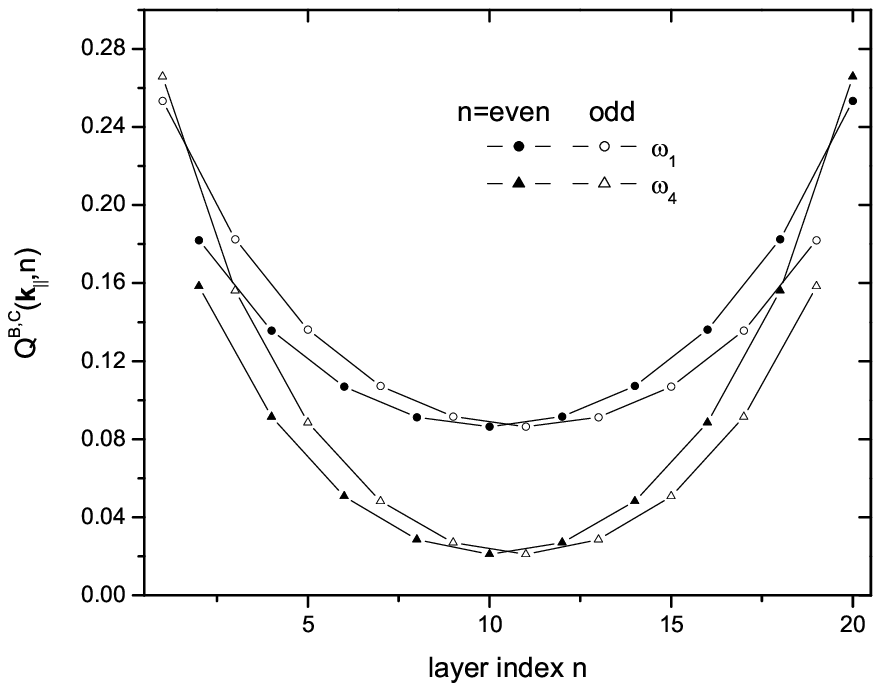}&\includegraphics[width=0.48\textwidth]{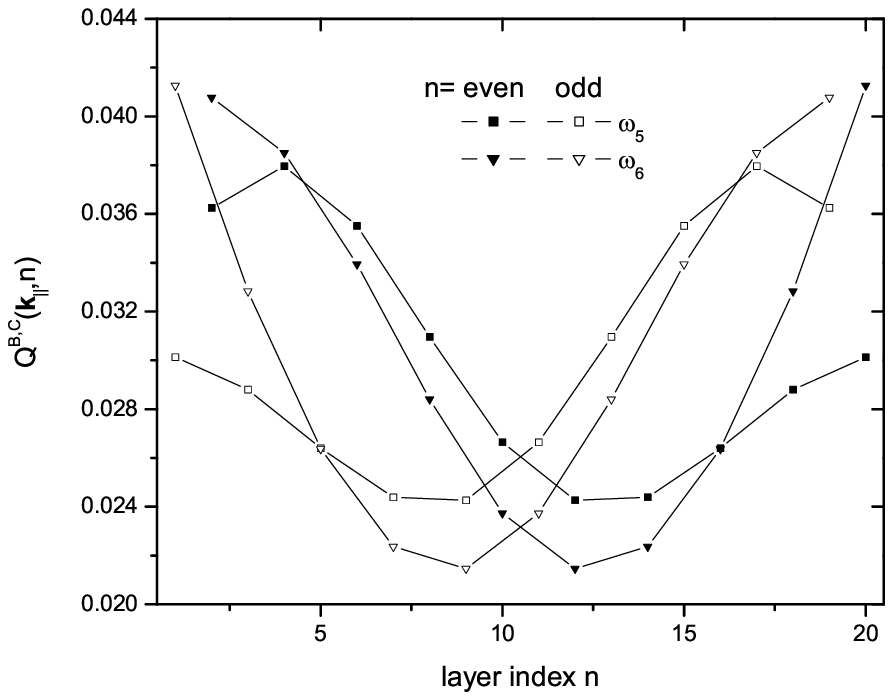}
\end{tabular}
\caption{\label{MSAA} Mean-squared amplitudes $Q^B(\mathbf{k_\parallel},n)=Q^C(\mathbf{k_\parallel},n)$ plotted against the layer index $n$ for the some of the low-energy $\mathbf{k_\parallel}=(2\pi/3a,0)$ modes defined in Fig. \ref{Q1DEA} for a quasi-1D film composed of $N=20$ layers.  The amplitudes $Q^B(\mathbf{k_\parallel},n)$ of the spin-wave modes $\omega_2$ and $\omega_3$ (not shown) are qualitatively similar to those of $\omega_1$.  All of the results are symmetric with respect to the center of the film because of the symmetric choice of anisotropy and exchange parameters at the surfaces.  The lines are guides to the eye.   The closed symbols define the amplitudes on even numbered layers whereas the open symbols are used for the odd numbered layers.  }
\end{figure}

The operator equation of motion method used in the previous sections may be easily extended to calculate Green functions which can be used to discuss the spectral intensities of the new spin-waves modes predicted in this work.   In order to further characterize the low-energy spin-wave modes in thin films with easy-axis anisotropy the mean-squared amplitude of the spin precession, defined for sublattice $i$ ($=A,B,C$) as $Q^i(\mathbf{k_\parallel},n)=\langle (S_{i,n}^x)^2+ (S_{i,n}^y)^2\rangle_{\mathbf{k_\parallel}}$, are considered.  For simplicity, we outline the calculation of $Q^A(\mathbf{k_\parallel},n)$.   This quantity can be written in terms of the transverse spin-correlation functions $\langle S_{A,n}^-(t) S_{A,n}^+(t') \rangle_{\mathbf{k_\parallel}}$ and $\langle S_{A,n}^+(t) S_{A,n}^-(t') \rangle_{\mathbf{k_\parallel}}$ evaluated at equal times ($t=t'$).  The spectral representation $\xi_n(\mathbf{k_\parallel},E)$ of the transverse spin-correlation function $\langle S_{A,n}^-(t) S_{A,n}^+(t') \rangle_{\mathbf{k_\parallel}}$ defined as 

\begin{eqnarray}
\langle S_{A,n}^-(t) S_{A,n}^+(t') \rangle_{\mathbf{k_\parallel}}=\int_{-\infty}^{\infty} \xi_n(\mathbf{k_\parallel},E) \exp[-iE(t-t')] \textrm{d}E,
\end{eqnarray}

\noindent is obtained from the imaginary part of the Green function ${G}^A_{n,n}(\mathbf{k_\parallel},E)=\ll S_{A,n}^+(k_\parallel);S_{A,n}^-(k_\parallel)\gg_E$ using the fluctuation dissipation theorem. \cite{Zubarev1}  The spectral function of $\langle S_{A,n}^+(t) S_{A,n}^-(t') \rangle_{\mathbf{k_\parallel}}$ may also be deduced from ${G}^A_{n,n}(\mathbf{k_\parallel},E)$. Some examples of the application of the Green function equation-of-motion method to magnetic thin films with a different ordering of the spins are given in Refs. [18-20].  In the low-temperature limit ($T\ll T_C$) the solution to the Green function ${G}^A_{n,n}(\mathbf{k_\parallel},E)$ may be written as 

\begin{eqnarray}
{G}^A_{n,n}(\mathbf{k_\parallel},E)=\frac{S}{\pi\det\mathbf{M}}[\textrm{adj } \mathbf{M}]_{6n-5,6n-5}\quad \textrm{for}\quad n=1,...,N
\end{eqnarray}

\noindent where $[\textrm{adj } \mathbf{M}]_{n,n}$ denotes the $n^{\textrm{th}}$ diagonal element of the adjoint of the matrix $\mathbf{M}$ defined in Eq. (\ref{eqmoeatf}).  Other diagonal elements of the adjoint of  $\mathbf{M}$  are used to obtain the Green functions ${G}^B_{n,n}(\mathbf{k_\parallel},E)$ and ${G}^C_{n,n}(\mathbf{k_\parallel},E)$.  The spin-wave energies obtained from $\det\mathbf{M}=0$ will correspond to the poles of the Green functions.   The mean-squared amplitude $Q^A(\mathbf{k_\parallel},n)$ (or integrated intensities) of a particular spin-wave mode is estimated by evaluating the area under the peak of the imaginary parts of ${G}^A_{n,n}(\mathbf{k_\parallel},E+i\epsilon)$, where the real and positive quantity $\epsilon$ is introduced phenomenologically to model an intrinsic damping or reciprocal lifetime.

Fig. \ref{MSAA} shows the mean squared amplitudes as a function of the layer index corresponding to some of the $\mathbf{k_\parallel}=(2\pi/3a,0)$ spin-wave modes shown in Fig. \ref{Q1DEA}.  The mean-squared amplitudes for sublattices $B$ and $C$ within any particular layer are equal.   Numerical results show that the spin-wave modes $\omega_1,...,\omega_4$ contribute mostly to the spin fluctuations on sublattices $B$ and $C$.  These modes are localized near the surfaces of the film and are characterized with amplitudes which decay with distance from the surfaces.  The amplitudes $Q^A(\mathbf{k_\parallel},n)$ (not shown) of the spin-wave modes $\omega_1,...,\omega_4$ are negligible.      The differences in amplitudes between odd and even layers is a result of the AF ordering of the spin chains.   The spin-wave modes $\omega_5$ and $\omega_6$ contribute much less to the fluctuations on sublattices $B$ or $C$ compared with modes $\omega_1,...,\omega_4$.  These higher energy modes are characterized with amplitudes that vary in a wavelike fashion across the thickness of the film on sublattices $B$ and $C$.   However, numerical results show that the spin-wave modes $\omega_5$ and $\omega_6$ are localized on sublattice $A$ sites that are near the surfaces the film.   Higher energy modes are characterized with amplitudes that vary in a wavelike fashion on all three sublattices.

\begin{figure}
\centering
\begin{tabular}{cc}
\includegraphics[width=0.48\textwidth]{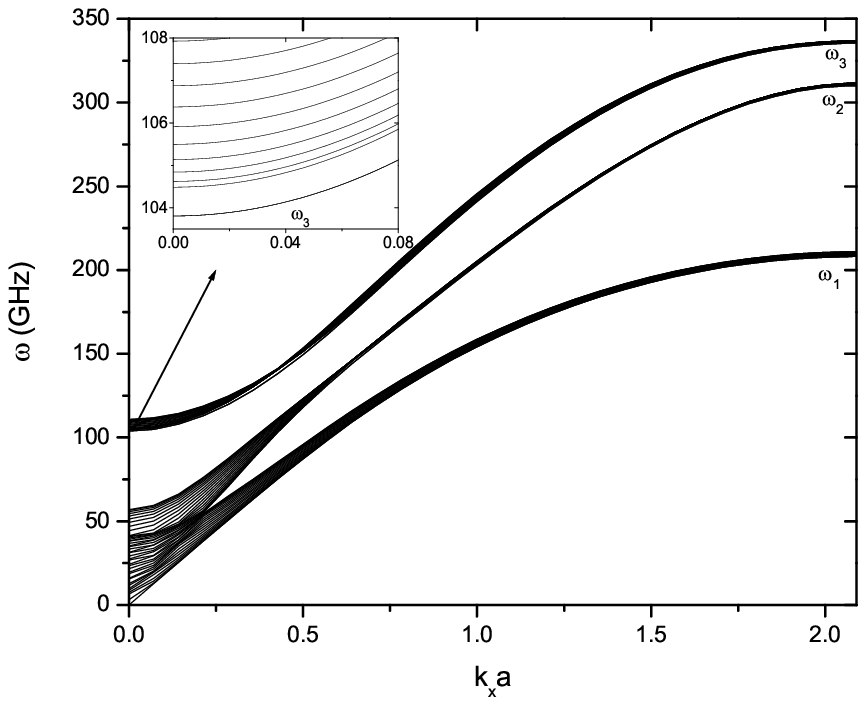}&\includegraphics[width=0.48\textwidth]{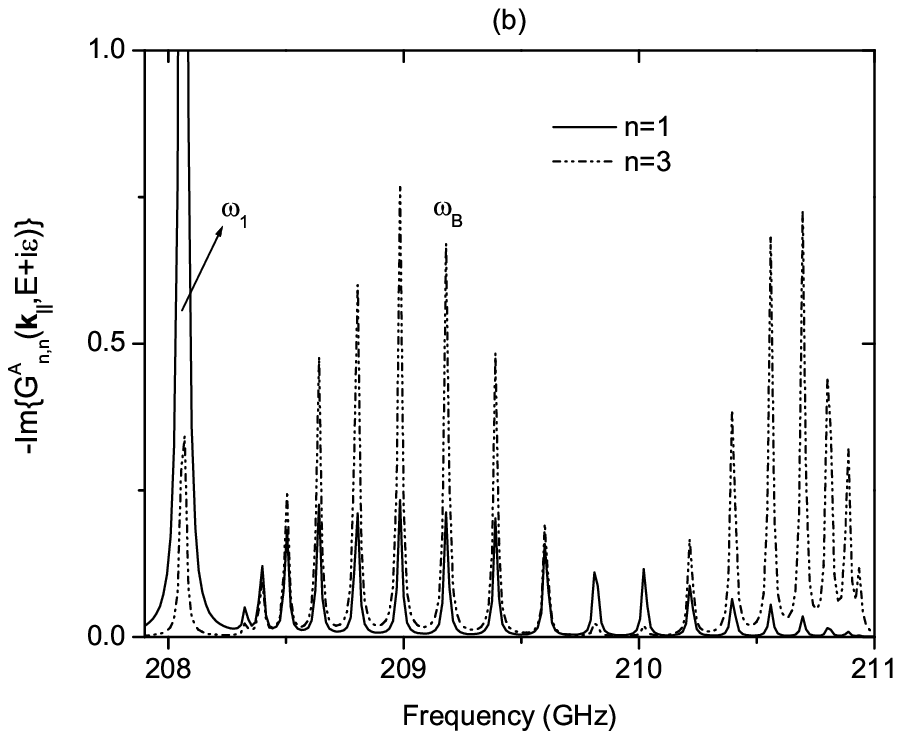}\\
\includegraphics[width=0.48\textwidth]{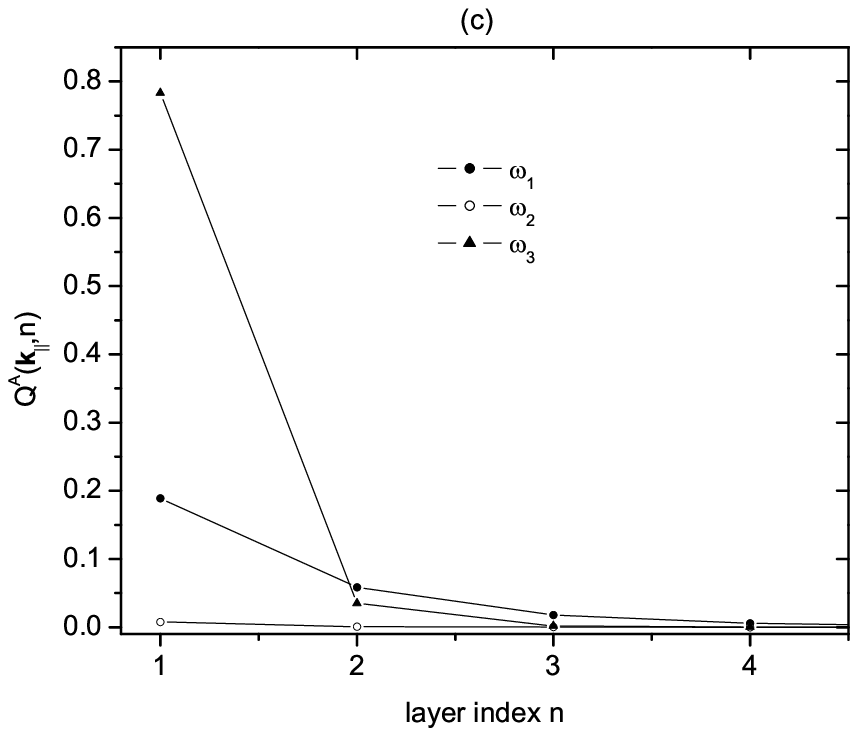}&\includegraphics[width=0.48\textwidth]{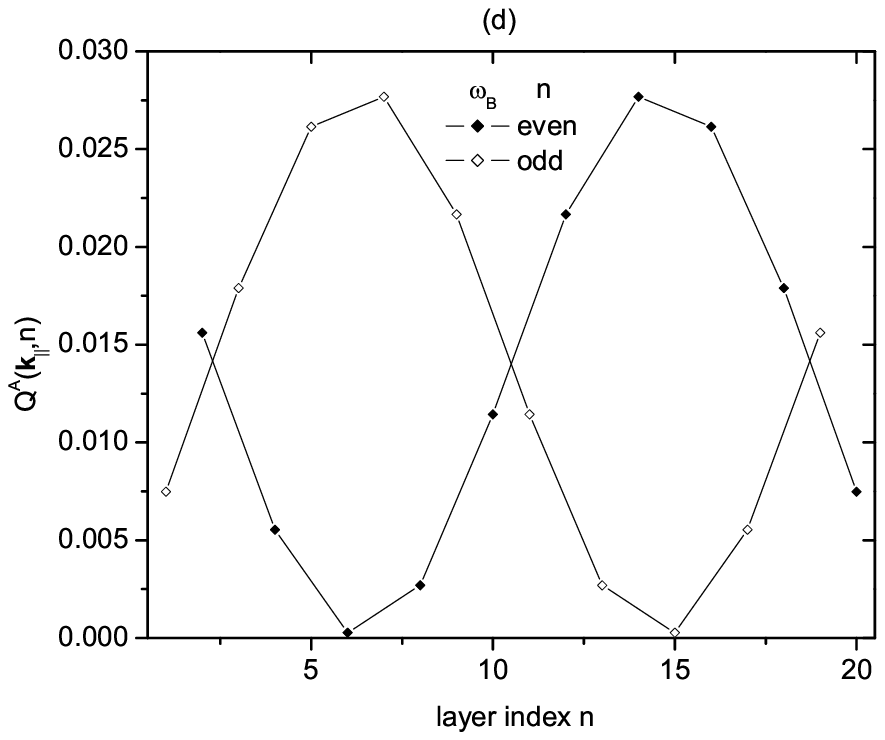}\\
\end{tabular}
\caption{ \label{q2D} (a): Spin-wave dispersion relation for a $S=1$ easy-axis quasi-2D film composed of 20 layers with uniform parameters where we set $D'=-10$ GHz, $J'=1.0$ GHz, $J=100.0$ GHz, and $\sigma=\sigma'=0.0$.  The spin-wave modes are split into 3 groups each containing 20 branches.  The labels $\omega_1$, $\omega_2$, $\omega_3$ refer to the lowest energy branches in each group.  The inset focuses on the small wavevector region and shows the splitting of the $\omega_3$ branch from the other modes in the group.    (b): Sublattice A spectral representation ($-\textrm{Im} \{{G}^A_{n,n}(\mathbf{k_\parallel},E+i\epsilon)\}$ with $\mathbf{k_\parallel}=(2\pi/3a,0)$ and $\epsilon=0.01$ GHz) versus frequency in layers $n=1$ (solid line) and $n=3$ (dashed line). (c): The amplitude $Q^A(\mathbf{k_\parallel},n)$ (integrated intensity) of the zone-edge surface modes $\omega_1$, $\omega_2$, $\omega_3$ as a function of the layer index.     Results are symmetric about the center of the film and only results from the first 4 layers are shown. Lines are guides to the eye and have no physical meaning. (d): The mean-squared amplitude $Q^A(\mathbf{k_\parallel},n)$ for the bulk excitation $\omega_B$ defined in Fig. \ref{q2D}b versus layer index.   Open and close symbols are used for odd and even layers, respectively.} 
\end{figure}

\begin{figure}
\centering
\begin{tabular}{cc}
\includegraphics[width=0.48\textwidth]{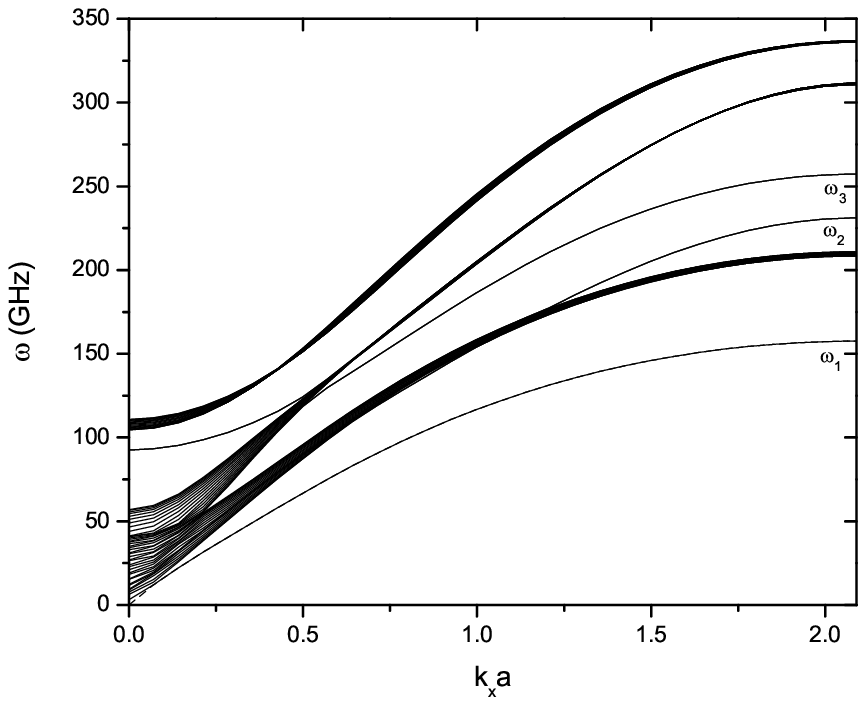}&\includegraphics[width=0.48\textwidth]{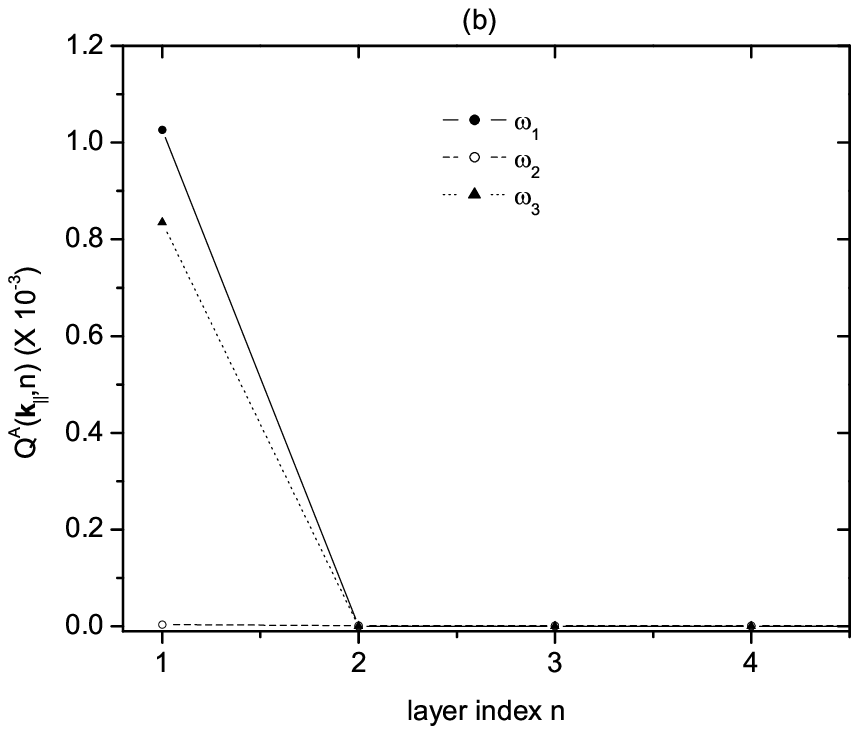}
\end{tabular}
\caption{ \label{q2D2} (a): As in Fig. \ref{q2D} except with modified exchange parameters at the surface of the film.  Results correspond to $J_1=J_N=0.9J$.  The labels $\omega_1$, $\omega_2$, $\omega_3$ refer to the 3 surface modes.   (b): Sublattice A mean-square amplitude evaluated at $\mathbf{k_\parallel}=(2\pi/3a,0)$ for the surface modes as a function of layer index showing the first 4 layers only.   The amplitudes are $\sim 3$ orders of magnitude smaller than those in Fig. \ref{q2D}c.}
\end{figure}

Fig. \ref{q2D} shows illustrative results for a $S=1$ quasi-2D film composed of $N=20$ layers where we set $D'=-10$ GHz, $J'=1.0$ GHz, $J=100.0$ GHz $\sigma=\sigma'=0.0$.   Exchange and anisotropy parameters at the surfaces are set equal to their corresponding bulk values.  For each value of the in-plane wavevector $\mathbf{k_\parallel}$ the solution of $\det\mathbf{M}=0$ yields a total of $6N$ spin-wave modes.  All of the solutions are degenerate in magnitude and the dispersion relation (showing positive solutions only) is illustrated in Fig. \ref{q2D}A.  The modes are split into 3 distinct groups each containing a total of 20 branches.  Qualitatively similar results (not shown) are obtained for a quasi-2D system with easy-plane anisotropy.   The main effect of the easy-axis anisotropy is to remove the degeneracies between some of the spin modes.   This is a consequence of the reduced symmetry in systems characterized with easy-axis anisotropy compared with systems with easy-plane anisotropy.  Discussions of the symmetry properties of the bulk spin-wave modes can be found in Ref. [13].   The lowest energy branch in each group is labelled $\omega_1$, $\omega_2$, and,  $\omega_3$.  In quasi-2D systems these modes are two-fold degenerate and each mode represents a localized excitation at the surfaces of the film.  The 18 other spin-waves branches within each group appear in the effective bulk continuum (not shown) obtained for the infinite system.   The inset in Fig. \ref{q2D}a shows the splitting in frequency (at small wavevectors) of the surface branch labelled $\omega_3$ from the rest of the modes in the group.   In Fig. \ref{q2D}b results are shown for the spectral representation obtained from $-\textrm{Im} \{{G}^A_{n,n}(\mathbf{k_\parallel},E+i\epsilon)\}$ for the 20 low-lying $\mathbf{k_\parallel}=(2\pi/3a,0)$ spin-wave excitations around $\omega_1$.   The solid and dashed lines show results for the spectral representation in layers $n=1$ and $n=3$, respectively.   The intensity of the spin-wave mode $\omega_1$ is largest in the surface layers $(n=1$ or $n=N)$ and decays extremely rapidly with distance away from the surfaces.   In Fig. \ref{q2D}c we show the amplitude $Q^A(\mathbf{k_\parallel},n)$ (integrated intensity) of the surface modes as a function of the distance away from the surface.   The results are symmetric about the center of the film and only the first 4 layers are shown.   Qualitatively similar behavior is obtained for the amplitudes $Q^B(\mathbf{k_\parallel},n)$  $(=Q^C(\mathbf{k_\parallel},n)$). The surface modes have negligible spectral intensities for any layer $4\le n\le 16$.  
Fig \ref{q2D}d shows results for the amplitude as a function of the layer index for the bulk mode labelled $\omega_B$ of Fig. \ref{q2D}b.   The modes appearing within the effective bulk continuum have amplitudes that vary in a wavelike fashion across the thickness of the film.  The intensities of the bulk excitations at the surfaces is strongly dependent on wavevector and the assumed anisotropy and exchange parameters at the surfaces of the film.

  Some effects of modified values of the surface exchange parameters on the spin-wave dispersion relation and the amplitude of the modes are illustrated in Fig. \ref{q2D2} for a quasi-2D system.  The results are obtained using the same parameters as in Fig. \ref{q2D}a except the exchange parameters at the surface are set as $J_1=J_N=0.9J$.   The localized surface modes labelled $\omega_1$, $\omega_2$, and,  $\omega_3$ are shifted to lower frequencies when the surface exchange parameters are less than the bulk values.  Fig. \ref{q2D2}b shows that the modes $\omega_1$, $\omega_2$, and  $\omega_3$, are localized near the surfaces of the film and are characterized with amplitudes that are orders of magnitude smaller than the case with uniform parameters throughout the thickness of the film.  For a symmetric film with $J_1=J_N$ and $D_1=D_N$ the surface modes are two-fold degenerate.   However, the degeneracy is lifted when the surfaces are asymmetric.

%%%%%%%%%%%%%%%%%%%%%%%%%%%%%%%%%%%%%%%%%%%%%%%%%%%%%%%%%%%%%%%%%%%%%%%%%%%%%%%%%%%%%%%%%%%%%%%%%%%%%%%%%%%%%%%%%%%
\section{Conclusions}\label{Conclusions}
%%%%%%%%%%%%%%%%%%%%%%%%%%%%%%%%%%%%%%%%%%%%%%%%%%%%%%%%%%%%%%%%%%%%%%%%%%%%%%%%%%%%%%%%%%%%%%%%%%%%%%%%%%%%%%%%%%%%
The results presented above outline the application of an operator formalism to calculate of surface and thin-film effects on the spin dynamics in AF coupled geometrically frustrated triangular layers. This serves to extend our previously published work on F coupled films to the physically more relevant case applicable to a large variety of materials.   Through the use of illustrative numerical calculations differences in linear spin excitations for bulk, semi-infinite and thin-film systems having either easy-plane or easy-axis anisotropy and quasi-1D or quasi-2D exchange couplings are highlighted.

    A number of features of these results are of interest.  In contrast with the case of F coupled films, bulk and surface mode excitations remain well separated in the case of AF coupled layers even in the case where surface exchange and anisotropy parameters are identical to bulk values.  Differences also arise between these two types of interlayer couplings in the case of thin films.   In the case of AF coupling, the results are dependent on whether there are an even or odd number of layers corresponding to having compensated or un-compensated total moments, respectively.  Most of our knowledge regarding the sign of exchange interactions is based on rules which may not be entirely applicable to surface and thin-film environments.\cite{Schneider1}  Examination of the spin excitation spectrum in view of these findings offers a possible tool to distinguish the type of interlayer coupling.  The present work also features a calculation of mode amplitudes.  These exhibit a variety dependences through the thickness of thin films dependent on the mode in question and if the system is quasi-1D or quasi-2D.
      
    These results generally are applicable to a wide variety of AF coupled layered triangular materials.  Among the ABX$_3$ compounds, CsMnBr$_3$ and CsVCl$_3$ are notable examples of quasi-1D materials with planar anisotropy.\cite{Oyedele1}  Bulk excitations have been well studied in prototypical axial quasi-1D hexagonal materials CsNiCl$_3$ and CsCoX$_3$.\cite{Nagler1}  Examples of quasi-2D compounds with well established bulk magnetic properties include axial CuFeO$_2$ \cite{Kimura1} and planar VX$_2$ and LiCrS$_2$.\cite{Laar1}  Most of the RMnO$_3$ materials exhibit planar anisotropy although characterization of thin films in this regard is ongoing.\cite{Fiebig1}   In ultrathin films surface effects dominate and spin reorientation transitions can occur.\cite{Li1}  A number of possible experimental techniques to examine long wavelength surface modes are summarized in our previous work.\cite{Meloche1}   Progress using inelastic neutron scattering techniques to probe surface excitation in multilayer systems has been reported.\cite{Schreyer1}  Potential extensions of the present calculations might be to include dipolar effects that can be important at surfaces even in AF systems.  This may require the utilization of purely numerical algorithms such as those based on classical equations of motion.\cite{Chaudhury1}

\begin{acknowledgments}
The work was partially supported by the Natural Sciences and Engineering Research Council of Canada (NSERC).
\end{acknowledgments}

\appendix
\section*{Appendix A}

The definition of matrices $\mathbf{A}_n$ and $\boldsymbol{\tau}_{n,n'}$ in Eq. (\ref{eqmoeatf}) for a stacked triangular antiferromagnetic thin film with easy-axis anisotropy.

\begin{eqnarray}\label{tfele}
\mathbf{A}_n=\left(\begin{array}{ccc}
\tilde{A}_n&\tilde{B}_n&\tilde{B}_n^*\\
\tilde{B}_n^*&\tilde{C}_n&\tilde{D}_n\\
\tilde{B}_n&\tilde{D}_n^*&\tilde{C}_n\\
\end{array}\right)\quad \textrm{,}\quad\boldsymbol{\tau}_{n,n'}=\left(\begin{array}{ccc}
{\tau}_{n,n'}&0&0\\
0&{\lambda}_{n,n'}&0\\
0&0&{\lambda}_{n,n'}\\
\end{array}\right)
\end{eqnarray}
\noindent where

\begin{eqnarray}
\tilde{A}_n&=&\left(\begin{array}{cc}
E+\Omega_n&0\\
0&E-\Omega_n
\end{array}\right);\quad \tilde{B}_n=-SJ(\mathbf{k}_\parallel)/2\left(\begin{array}{rr}
 c_{1,n}^+&c_{1,n}^- \\
-c_{1,n}^- & -c_{1,n}^+
\end{array}\right);\quad \tilde{C}_n=\left(\begin{array}{cc}
E+ \alpha_n& \delta_n\\
-\delta_n & E-\alpha_n
\end{array}\right);\nonumber\\ \tilde{D}_n&=&-SJ(\mathbf{k}_\parallel)/2\left(\begin{array}{rr}
 c_{2,n}^+&c_{2,n}^- \\
-c_{2,n}^- & -c_{2,n}^+
\end{array}\right);\quad {\tau}_{n,n'}=SJ'\left(\begin{array}{rr}
0 &1\\
-1&0\end{array}\right);\\
\quad {\lambda}_{n,n'}&=&SJ'/2\left(\begin{array}{cc}
\sigma'\sin\theta_n\sin\theta_{n'} & 2\cos(\theta_n-\theta_{n'})+\sigma'\sin\theta_n\sin\theta_{n'}\\
-2\cos(\theta_n-\theta_{n'})+\sigma'\sin\theta_n\sin\theta_{n'}& -\sigma'\sin\theta_n\sin\theta_{n'}\end {array}\right)\nonumber\\
\end{eqnarray}

\noindent with matrix elements defined as
\begin{eqnarray}
\Omega_n&=&2S(1+\sigma)\cos\theta_n J(0)+2SD'-S(1+\sigma')J'[2-\delta_{1,n}-\delta_{n,N}]\nonumber\\
\alpha_n&=&S\big((1+\sigma)(\cos\theta_n+\cos^2\theta_n)-\sin^2\theta_n\big)J(0)-S(1-3\cos^2\theta_n)D'\nonumber\\
&-&SJ'\big([\cos(\theta_n-\theta_{n-1})+\sigma'\cos\theta_n\cos\theta_{n-1}](1-\delta_{1,n})\nonumber\\
&+&[\cos(\theta_n-\theta_{n+1})+\sigma'\cos\theta_n\cos\theta_{n+1}](1-\delta_{n,N})\big)\nonumber\\
\delta_n&=&-S D'\sin^2\theta_n\nonumber\\
c_{1,n}^\pm&=&\cos\theta_n\pm1\nonumber\\
c_{2,n}^\pm&=&\cos^2\theta_n-(1+\sigma)\sin^2\theta_n\pm 1\nonumber\\
D'_n&=&[1-(2S)^{-1}]D_n.
\end{eqnarray}

\end{document}